\documentclass[manuscript]{aastex}

\shorttitle{ASTE CO($J=3-2$) observation of M~83}
\shortauthors{Muraoka et al.}

\begin{document}


\title{ASTE CO(3-2) Mapping toward the Whole Optical Disk of M~83:\\
Properties of Inter-arm GMAs}


\author{Kazuyuki Muraoka\altaffilmark{1,2}, Kotaro Kohno\altaffilmark{3,4},
Tomoka Tosaki\altaffilmark{5}, Nario Kuno\altaffilmark{1},\\
Kouichiro Nakanishi\altaffilmark{6}, Kazuo Sorai\altaffilmark{7},
Tsuyoshi Sawada\altaffilmark{6,8}, Kunihiko Tanaka\altaffilmark{9},\\
Toshihiro Handa\altaffilmark{3}, Masayuki Fukuhara\altaffilmark{1},
Hajime Ezawa\altaffilmark{6}, and Ryohei Kawabe\altaffilmark{1}}
\email{kmuraoka@p.s.osakafu-u.ac.jp}


\altaffiltext{1}{Nobeyama Radio Observatory, Minamimaki, Minamisaku, Nagano 384-1305}
\altaffiltext{2}{Department of Physical Science, Osaka Prefecture University, Gakuen 1-1, Sakai, Osaka 599-8531}
\altaffiltext{3}{Institute of Astronomy, The University of Tokyo, 2-21-1 Osawa, Mitaka, Tokyo 181-0015}
\altaffiltext{4}{Research Center for the Early Universe, University of Tokyo, 7-3-1 Hongo, Bunkyo, Tokyo 113-0033}
\altaffiltext{5}{Department of Geoscience, Joetsu University of Education, Joetsu, Niigata 943-8512}
\altaffiltext{6}{National Astronomical Observatory of Japan, 2-21-1 Osawa, Mitaka, Tokyo 181-8588}
\altaffiltext{7}{Division of Physics, Graduate School of Science, Hokkaido University, Sapporo 060-0810}
\altaffiltext{8}{Joint ALMA Office, El Golf 40, Piso 18, Las Condes, Santiago, Chile}
\altaffiltext{9}{Department of Physics, Faculty of Science and Technology, Keio Univ., 3-14-1 Hiyoshi, Yokohama, Kanagawa, Japan}


\begin{abstract}
We present a new on-the-fly (OTF) mapping of CO($J=3-2$) line emission
with the Atacama Submillimeter Telescope Experiment (ASTE)
toward the $8' \times 8'$ (or 10.5 $\times$ 10.5 kpc at the distance of 4.5 Mpc) region of
the nearby barred spiral galaxy M~83 at an effective resolution of 25$^{\prime \prime}$.
Due to its very high sensitivity, our CO($J=3-2$) map can depict not only spiral arm structures but also
spur-like substructures extended in inter-arm regions.
This spur-like substructures in CO($J=3-2$) emission are well
coincident with the distribution of massive star forming regions traced by H$\alpha$ luminosity
and Spitzer/IRAC 8 $\mu$m emission.
We have identified 54 CO($J=3-2$) clumps as Giant Molecular-cloud Associations (GMAs)
employing the CLUMPFIND algorithm, and have obtained their sizes, velocity dispersions,
virial masses, and CO luminosity masses.
We found that the virial parameter $\alpha$, which is defined as the ratio of the virial mass
to the CO luminosity mass, is almost unity for GMAs in spiral arms,
whereas there exist some GMAs whose $\alpha$ are 3 -- 10 in the inter-arm region.
We found that GMAs with higher $\alpha$ tend not to be
associated with massive star forming regions, while other virialized GMAs are.
Since $\alpha$ mainly depends on velocity dispersion of the GMA,
we suppose the onset of star formation in these unvirialized GMAs with higher $\alpha$ are suppressed
by an increase in internal velocity dispersions of Giant Molecular Clouds within these GMAs
due to shear motion.
\end{abstract}





\section{Introduction}

The dense molecular medium is an indispensable
component of our understanding of the star formation law in galaxies.
This is because stars are formed from dense molecular cores,
not from the diffuse envelopes of Giant Molecular Clouds (GMCs).
In order to understand the global distribution of a dense
molecular medium in galaxies, we have conducted an extragalactic
CO($J=3-2$) imaging survey of nearby spiral galaxies,
ADIoS (ASTE Dense gas Imaging of Spiral galaxies).
The submillimeter-wave CO($J=3-2$) line emission is a tracer of dense gas,
because its Einstein A coefficient is proportional to $\nu^3$; therefore,
the critical density of CO($J=3-2$) line emission is higher than that of CO($J=1-0$) by a factor of $3^3$,
i.e., $n_{\rm H_2}$ $\sim$ $10^4$ cm$^{-3}$.
The current sample galaxies of the ADIoS project includes a GMA in M~31 \citep{tosaki2007a}, 
NGC~604 in M~33 \citep{tosaki2007b}, M~83 \citep{muraoka2007},
and NGC~986 \citep[and see a summary therein]{kohno2008}.
Several authors also report large-area CO($J=3-2$) mappings covering disk regions of galaxies.
For example, \cite{minamidani2008} performed CO($J=3-2$) line emission observations of 6 GMCs
in the Large Magellanic Cloud (LMC). They identified 32 molecular clumps in the GMCs,
and obtained their physical properties: kinetic temperature and density of molecular gas.
\cite{walsh2002} carried out the CO($J=3-2$) observation toward $\sim 6^{\prime} \times 4^{\prime}$ of NGC~6946,
and they found that the arm/inter-arm contrast of CO($J=3-2$) is significantly larger than
that of CO($J=1-0$). \cite{wilson2009} performed the CO($J=3-2$) mapping of some members of Virgo Clusters
(NGC~4254, NGC~4321, and NGC~4569), and they examined a large-scale gradient in the gas properties
traced by CO($J=3-2$)/CO($J=1-0$) intensity ratio.

The Atacama Submillimeter Telescope Experiment \citep[ASTE:][]{ezawa2004, ezawa2008},
a new project to operate a 10 m telescope in the Atacama Desert of northern Chile,
provides us with an ideal opportunity to generate high-sensitivity and large-scale maps of CO($J=3-2$) line emission,
because of the high beam efficiency of the telescope, a low system noise temperature
due to the excellent receiver system and atmospheric conditions at the site,
and its efficient on-the-fly (OTF) mapping capability \citep{sawada2008}.
The excellent performance in CO mapping enables us to detect and
resolve a few 100 pc scale structures of molecular gas in nearby galaxy disks.
Molecular gas structures on this scale are often referred to as Giant Molecular-cloud Associations (GMAs)
\citep{vogel1988, rand1990}, whose typical masses are $\sim 10^7 M_{\odot}$.
Observational studies of GMAs in galaxies provide us with invaluable clues to the physics
that governs large-scale star formation in the disk regions of galaxies
\citep[e.g.][]{kuno1995, tosaki2003, sakamoto1996, wong2002, lundgren2004}.
Of course, it is essential to investigate physical parameters of GMAs themselves,
i.e., their distributions, sizes, velocity dispersions, masses, and luminosities
in order to understand their origin and nature.

Studies of GMA properties in galaxies have been mainly performed
toward the grand-design spiral galaxy M~51.
For example, \cite{rand1990} identified 26 GMAs in the spiral arm and the inter-arm of M~51
at a spatial resolution of 500 pc $\times$ 350 pc.
They found that most GMAs in the spiral arm appeared to be bound,
whereas those in the inter-arm appeared to be unbound,
although they could only give the upper limit of virial masses for 14 GMAs
due to the limitation of sensitivity.
\cite{rand1993} studied the issue of GMA formation and destruction.
The author concluded that both gravitational instability and collisional agglomeration
of molecular gas were viable mechanisms for producing bound GMAs in the spiral arm.
In addition, \cite{aalto1999} obtained the higher-quality CO($J=1-0$) image of M~51.
The authors discussed the velocity gradients across the spiral arms in detail,
and they found that the velocity gradients generally become steeper between GMAs
than within them.
However, these studies are limited to the relatively narrow,
central $2^{\prime}.5$ -- $3^{\prime}$ region of M~51.
So they could not obtain a spatially unbiased GMA sample covering the whole galaxy disk.
Moreover, there are few studies of GMA properties toward other galaxies.
Therefore, high angular resolution and high sensitivity CO images
covering whole galaxy disks are required in order to obtain a sufficient number of GMA samples
and to study not only properties of GMAs but also their relationships with star formation.

In this paper, we present new CO($J=3-2$) images of the nearby barred spiral galaxy M~83
using the ASTE, employing OTF mapping mode.
M~83 is a nearby, face-on, barred, grand-design spiral galaxy hosting an intense starburst at its center.
The distance to M~83 is estimated to be 4.5 Mpc \citep{thim2003};
therefore, 1$^{\prime \prime}$ corresponds to 22 pc.
The inclination of M 83 is 24$^{\circ}$ \citep{comte1981}.
Its proximity and face-on view enable us to resolve not only spiral arm structures
but also each GMA within these structures even by single-dish observations at millimeter or submillimeter wavelength.
In addition, substructures diverging from the bar-ends and spiral arms are already found in M~83.
Figure~1 shows the Spitzer/IRAC 8 $\mu$m image, depicting not only clear-cut spiral arm structures
but also a number of capillary substructures stretching from spiral arms.
These substructures are often referred to as ``spurs'' \citep[and references therein]{wada2008}.
In this study, we divided the M~83 disk into two regions; the spiral arm region and the inter-arm region.
We refer to the continuous and thick structures seamlessly linked from the bar in the 8 $\mu$m image as the ``spiral arm'',
and refer to the other region, including spurs, as the ``inter-arm''. 
If we detect molecular gas associated with spurs,
we can study details of the difference in GMA properties between spiral arm and inter-arm regions.
Until recently, CO($J=3-2$) images of M~83 were limited to the nuclear region and the bar
\citep{petitpas1998, israel2001, dumke2001, sakamoto2004, bayet2006}.
\citet{muraoka2007} presented wider CO($J=3-2$) image including inner spiral arms,
but they could not detect CO($J=3-2$) line emission in inter-arm regions and not cover outer spiral arms.
In addition, their CO($J=3-2$) map is not sufficiently sensitive to resolve and identify each GMA in the M~83 disk.

The goals of this paper are:
(1) to depict the global distribution of CO($J=3-2$) line emission which is wider and more sensitive than 
previous CO($J=3-2$) image \citep{muraoka2007},
(2) to identify GMAs throughout the whole M~83 disk, and to obtain their physical parameters
(sizes, velocity dispersions, luminosities, and masses),
and (3) to investigate differences between GMAs on spiral arms and those in inter-arm regions
comparing their physical parameters, and finally to study the origin and the detailed nature of inter-arm GMAs.

Further analysis of the quantitative relationship between CO($J=3-2$)/CO($J=1-0$) ratios
and star formation will be presented in forthcoming papers.

\section{Observations and Data Reduction}

CO($J=3-2$) line emission observations of M 83 were performed
using the ASTE from May 15 -- 21, 2008.
The total time for the observations was 42 hours.
The size of the CO($J=3-2$) map is about $8' \times 8'$ (10.5 $\times$ 10.5 kpc), including the whole optical disk.
The coverage of $8' \times 8'$ is the largest of all the CO($J=3-2$) mappings towards external galaxies.
The mapped region is indicated in figure~2.
The full-half power beam width (HPBW) of the ASTE 10 m dish is 22$^{\prime \prime}$ at this frequency;
this corresponds to 480 pc at a distance of 4.5 Mpc.

We used a waveguide-type sideband-separating SIS mixer receiver for the single side band (SSB) operation, CATS345 \citep{ezawa2008, inoue2008}.
The image rejection ratio at 345 GHz was estimated to be $\sim$ 10 dB.
The backend was a digital autocorrelator system, MAC \citep{sorai2000}, which comprises four banks of a 512 MHz
wide spectrometer with 1024 spectral channels each.
This arrangement provided a velocity coverage of 440 km s$^{-1}$ with a velocity interval of 0.43 km s$^{-1}$.
The observations were made remotely from the ASTE operation room at the Nobeyama Radio Observatory (NRO)
using a network observation system N-COSMOS3 developed at NAOJ \citep{kamazaki2005}.
Since the observations were carried out in excellent atmospheric conditions
(the opacity of 220~GHz was always lower than 0.05),
the system noise temperature was typically 200 K (in SSB).

OTF mapping was performed along two different directions
(i.e., scans along the RA and Dec directions), and these two
data sets were co-added by the Basket-weave method \citep{emerson1988}
in order to remove any effects of scanning noise. At the beginning of each OTF scan,
an off-source position, which was 10$^{\prime}$ offset in the R.A. direction from the map center taken
at $\alpha = 13^h 37^m 00^s .48$, $\delta = -29^{\circ} 51^{\prime} 56^{\prime \prime}.5$ (J2000),
was observed to subtract sky emission.

We observed the CO($J=3-2$) line emission of a carbon star II Lup every 2 hours in order to check
the stabilities of the pointing accuracy, which was found to be better than $\sim$ 2$^{\prime \prime}$ r.m.s.
In addition, we observed the CO($J=3-2$) line emission of CW Leo and W~28 to obtain the main beam efficiency $\eta_{\rm MB}$
every day. We compared our CO($J=3-2$) spectra with those obtained by CSO observations \citep{wang1994},
and $\eta_{\rm MB}$ was estimated to be 0.54 -- 0.60 during the observing runs.
The absolute error of the CO($J=3-2$) amplitude scale was about $\pm$ 20\%,
mainly due to variations in $\eta_{\rm MB}$ and the image rejection ratio of the CATS345.

The data reduction was made using the software package NOSTAR, which comprises tools for OTF data analysis,
developed by NAOJ \citep{sawada2008}. The raw data were regridded to 7$^{\prime \prime}$.5 per pixel,
giving an effective spatial resolution of approximately 25$^{\prime \prime}$ (or 550 pc).
Linear baselines were subtracted from the spectra.
We binned the adjacent channels to a velocity resolution of 5 km s$^{-1}$ for the CO($J=3-2$) spectra.
Going through these procedures, a 3D data cube was created.
The resultant r.m.s.\ noise level (1 $\sigma$) was typically in the range of 25 to 30 mK in the $T_{\rm MB}$ scale
at a beam size of 25$^{\prime \prime}$ (HPBW).
The observation parameters of M~83 are summarized in table~1.

\section{Results}

\subsection{CO($J=3-2$) Spectra}

Figure~3 shows spectra of the CO($J=3-2$) line emission of M~83.
In addition, figure~4 shows the spectra in the center, the eastern bar-end, and the western spiral arm,
where the peak temperature of the CO($J=3-2$) line emission exceeds 0.5~K.
We have detected significant CO($J=3-2$) line emission not only in the central region (peak temperature $>$ 1.0 K),
the bar (0.3 -- 0.5 K), the bar-ends (0.4 -- 0.8 K), and the spiral arms (0.2 -- 0.7 K)
but also in the inner inter-arm regions
(0.1 -- 0.2 K) which are surrounded by the central region and the western spiral arm.
In addition, we have detected substructures diverging from spiral arms in the CO($J=3-2$) line emission.
These substructures are molecular gases associated with spurs.
\cite{aalto1999} also detected such molecular spurs in M~51.

Velocity-integrated intensities, $I_{\rm CO(3-2)}$, were calculated from the CO($J=3-2$) line emission
within a velocity range of 380 to 620 km s$^{-1}$ using the NRAO AIPS task MOMNT.
At the map center, we found $I_{\rm CO(3-2)}$ = 147 $\pm$ 2 K km s$^{-1}$ at a resolution of 25$^{\prime \prime}$.
This $I_{\rm CO(3-2)}$ value seems in excellent agreement with that of the previous observation \citep{muraoka2007}, 161 $\pm$ 4 K km s$^{-1}$,
at a resolution of 22$^{\prime \prime}$ if we consider the small difference in the spatial resolution.

In addition, we have calculated the luminosity of CO($J=3-2$), $L^{\prime}_{\rm CO(3-2)}$,
over the observed $8^{\prime} \times 8^{\prime}$ region.
We found $L^{\prime}_{\rm CO(3-2)}$ = 5.7 $\times$ $10^8$ K km s$^{-1}$ pc$^2$,
which is slightly greater than the previously obtained value, 5.1 $\times$ $10^8$ K km s$^{-1}$ pc$^2$ ($r$ $<$ 3.5 kpc),
by \citet{muraoka2007}.
This is consistent with the fact that our new CO($J=3-2$) map covers a wider region of the M~83 disk than the previous one.
The derived $L^{\prime}_{\rm CO(3-2)}$ of M~83 is comparable with that of
the whole disk region of the prominent barred galaxy NGC~986, 5.4 $\times$ $10^8$ K km s$^{-1}$ pc$^2$ \citep{kohno2008}.
It should be noteworthy that the extended CO($J=3-2$) emission contributes a lot to the total luminosity;
the central ($r<30''$) CO($J=3-2$) luminosity is $9 \times 10^7$ K km s$^{-1}$ pc$^2$, just $\sim$ 1/6 of the total luminosity.
This result emphasizes that it is important to observe extended component in order to quantify the total luminosity
of CO even in a higher transition like $J=3-2$.
Map parameters and physical properties of M~83 are listed in table~2.

\subsection{CO($J=3-2$) Integrated Intensity Map and Velocity Field}

Figure~5a shows an integrated intensity map in CO($J=3-2$) line emission of M~83.
The 1 $\sigma$ noise level of this map is 1.0 K km s$^{-1}$.
Compared with the previous CO($J=3-2$) map obtained by \citet{muraoka2007},
spiral arms are more clearly depicted.
As seen in the spectra (figure~3), molecular substructures associated with spurs
also appear in the CO($J=3-2$) integrated intensity map.
Some of them seem bridges connecting inner bar-ends/spiral arms and outer arms.
Figure~5b displays a velocity field in the CO($J=3-2$) line.
A warp of velocity field is clearly seen in the western spiral arm,
suggesting the existence of streaming motion.

\subsection{Comparison with other wavelengths: H$\alpha$ and Near to Mid Infrared}

We found that the spur-like substructures in CO($J=3-2$) line emission are well spatially coincident
with the distribution of massive star forming regions traced
by H$\alpha$ luminosity and Spitzer/IRAC 8 $\mu$m emission as shown in figure~6.
These coincidences strongly suggest that the spur-like substructures appearing in CO($J=3-2$) line emission are
neither the noise nor the artifact produced during the data reduction but significant structures of M~83.
Good spatial correlation between CO($J=3-2$) line emission and massive star formation
in spiral arms of M~83 has already reported by \citet{muraoka2007},
but we newly found that this spatial correlation is true even in the inter-arm region.

In addition, we compared CO($J=3-2$) line intensities with H$\alpha$ luminosities 
at a resolution of $25^{\prime \prime}$ over the whole M~83 disk.
Here we employed calibrated H$\alpha$ luminosities using the Spitzer/MIPS 24 $\mu$m data
according to the procedure introduced by \cite{calzetti2007}.
We used an H$\alpha$ image obtained with the CTIO 1.5 m telescope \citep{meurer2006}
and an archival MIPS 24 $\mu$m image (P00059, George, Rieke, Starburst activity in nearby galaxies).
The formula to calibrate H$\alpha$ luminosity using 24 $\mu$m image is as follows \citep{calzetti2007}:
\begin{eqnarray}
L_{\rm H \alpha, cal} = L_{\rm H \alpha} + (0.031 \pm 0.006) L_{\rm 24 \mu m} \,\,\, {\rm erg} \,\, {\rm s^{-1}}
\end{eqnarray}
where $L_{\rm H \alpha, cal}$ and $L_{\rm H \alpha}$ means calibrated and observed H$\alpha$ luminosities, respectively.
$L_{\rm 24 \mu m}$ means the observed 24 $\mu$m luminosity.
Figure~7 shows the correlation between CO($J=3-2$) intensities and calibrated H$\alpha$ luminosities
over the whole M~83 disk. We found the index of this correlation to be 0.96 $\pm$ 0.12,
which is significantly lower than that of the standard Schmidt law using CO($J=1-0$) intensity, 1.4 \citep{kennicutt1998}.
This tendency for the index of ``CO($J=3-2$) Schmidt law'' to be lower than that of ``CO($J=1-0$) Schmidt law''
is also found by \cite{komugi2007} and \cite{iono2009}.
In addition, similar situations are also found in studies of HCN($J=1-0$) line emission, an excellent dense gas tracer.
\cite{gao2004} found the linear correlation between HCN($J=1-0$) luminosities and star formation rates
traced by infrared luminosities over three orders of magnitude. 
Moreover, numerical simulations also predict a linear correlation
between volume-averaged density of star formation rate and that of molecular gas \citep{wada2007}.

CO emission line from molecular clouds is generally optically thick,
but the observation beam contains many molecular clouds
in CO emission observation of galaxies.
Therefore, measurement of CO line intensity corresponds to 
counting up the number of molecular clouds.
This concept is well known as a ``mist'' model, proposed by \cite{solomon1987}.
In observations of CO($J=3-2$) line emission and HCN($J=1-0$) line emission
which trace a denser part of the molecular gas, measurements of their intensity
correspond to the number of star forming dense cores rather than
that of molecular clouds. Therefore, CO($J=3-2$) line intensity can trace
the inner portion of the molecular clouds, i.e., denser part of the molecular clouds,
even if the line is optically thick.
Therefore, the intensity is proportional to the amount of massive star formed in dense molecular clouds.

On the other hand, the fact that the index of ``CO($J=1-0$) Schmidt law'' is 1.4
means that the more massive molecular gas, and the higher star formation efficiency,
which is defined as star formation rate per unit gas mass.
This suggests the fraction of dense molecular gas become higher in massive molecular clouds
than in less massive molecular clouds.
It is unclear why there is such a tendency yet.
In order to study the index of Schmidt law more detail, we need both further analysis
using other spectral lines such as $^{13}$CO($J=1-0$) and HCN($J=1-0$)
and the progress of numerical study concerning the evolution of
interstellar matter and star formation.

In figure~5 and 6, one can find several local CO peaks in the spiral arm and isolated clump-like CO peaks in inter-arm regions,
associated with giant HII regions.
These clump-like structures whose sizes are typically a few $\times$ 100 pc correspond to GMAs,
which are literally ensembles of GMCs.
GMAs are important gaseous constituents of galaxies,
because the kpc-scale major structures such as the bar and the spiral arms
consist of multiple GMAs.

\section{Discussion}

Since our new CO($J=3-2$) data set covers a wider region (10.5 $\times$ 10.5 kpc)
and achieves a significantly low noise level (1 $\sigma$ $\sim$ 25 -- 30 mK in $T_{\rm MB}$),
we were able to find many individual CO($J=3-2$) clumps over the M~83 disk.
In order to identify each clump objectively and to obtain its physical parameters,
we employed the CLUMPFIND algorithm developed by \citet{williams1994}.

In this section, we discuss relationships among the size, velocity dispersion, and luminosity of each GMA
identified as a clump according to the CLUMPFIND algorithm.
Then, we study the properties of GMAs in the M~83 disk throughout the comparison of their virial masses
and CO luminosity masses.
Finally, we investigate the origin and the nature of GMAs in the inter-arm region.

\subsection{GMA Identification}

First, we ran the CLUMPFIND software setting a peak threshold of 4 $\sigma$ and a contour increment of 1 $\sigma$.
In this condition, we identified 99 clumps in the whole M~83 disk.
Then, we excluded clumps without adjacent pixels exceeding 3 $\sigma$ or 
adjacent channel exceeding 3 $\sigma$.
In addition, we excluded clumps whose diameters are smaller than effective spatial resolution
of our CO($J=3-2$) data ($25^{\prime \prime}$ $\sim$ 550 pc) 
after the deconvolution of clump size with 25$^{\prime \prime}$ beam,
according to the following formula,
\begin{eqnarray}
\theta_{\rm decon} = \sqrt[]{\theta_{\rm given}^2 - 25^2} \,\,\, .
\end{eqnarray}
Here, $\theta_{\rm decon}$ is a deconvolved clump size in arcsec unit,
and $\theta_{\rm given}$ is a clump size determined by the CLUMPFIND algorithm.
Finally, we identified 54 clumps in the M~83 disk ($r$ $>$ 1 kpc),
and obtained their radii, velocity dispersions, and CO($J=3-2$) luminosities.
Hereafter, we refer to these identified clumps simply as ``GMAs'' in this paper.
We show an example of the individual spectrum of an identified GMA in figure 8.
This shows a Gaussian shape (a fitted line is also displayed in the figure),
and the most of the identified clumps tend to have a similar shape.
This is consistent with the mist model by \cite{solomon1987}.
We obtained the total luminosity of all the identified GMAs to be 2.1 $\times$ $10^8$ K km s$^{-1}$ pc$^2$,
which corresponds to 44 \% of total luminosity of the M~83 disk, 4.8 $\times$ $10^8$ K km s$^{-1}$ pc$^2$.
The other 56\% luminosity may originate in diffuse components of molecular gas.

Note that we were unable to identify narrow line-width GMAs due to finite velocity resolution (5 km s$^{-1}$).
The lower limit of line width that we can detect is 10.3 km s$^{-1}$ 
(corresponding to the velocity dispersion $\sigma_v$ = 4.4 km s$^{-1}$)
for 4 $\sigma$ peak GMA and 8.2 km s$^{-1}$ ($\sigma_v$ = 3.5 km s$^{-1}$)
for 5 $\sigma$ peak GMAs if we assume the Gaussian line profile.
In this condition, the lower limit of CO intensity and CO luminosity are
1.2 K km s$^{-1}$ and 4.4 $\times 10^5$ K km s$^{-1}$ pc$^2$, respectively.
This corresponds to the lower limit of CO luminosity mass of 1.8 $\times 10^6$ $M_{\odot}$ (see subsection 4.3.2).

\subsection{GMA Distribution in M~83}
Figure~9 and 10 show positions of identified GMAs on the CO($J=3-2$) velocity channel maps
and on the Spitzer/IRAC 8 $\mu$m map, respectively.
GMAs are widely distributed over the M~83 disk.
We divided these GMAs into two groups according to the location: the spiral arm and the inter-arm.
We identified 27 GMAs in the spiral arm, and refer to these as ``on-arm GMAs''.
Meanwhile, we also identified 27 GMAs in the inter-arm, and refer to these as ``inter-arm GMAs''.

Obtained properties of identified GMAs are summarized in table 3 and 4.
Except for some GMAs, $\sigma_v$ of GMAs are mostly smaller than 10 km s$^{-1}$.
Any GMA with larger $\sigma_v$ exceeding 10 km s$^{-1}$ are distributed near the center and the bar.
Strong non-circular motion along the molecular bar \citep{sakamoto2004, muraoka2009}
may affect these GMAs, and consequently increase their velocity dispersions.

\subsection{Comparison between on-arm GMAs and inter-arm GMAs}
\subsubsection{CO($J=3-2$) luminosity $L^{\prime}_{CO(3-2)}$}

Comparing properties between on-arm GMAs and inter-arm GMAs,
we found the difference in the GMA radius $R$ and the CO($J=3-2$) luminosity $L^{\prime}_{CO(3-2)}$.
Figure~11 shows the histograms of $R$ and $L^{\prime}_{CO(3-2)}$ for on-arm and inter-arm GMAs.
We found that the averaged values of $R$ and $L^{\prime}_{\rm CO(3-2)}$
for on-arm GMAs are both greater than those for inter-arm GMAs.
In addition, there exist some on-arm GMAs whose $R$ are comparable to 1 kpc and
whose $L^{\prime}_{\rm CO(3-2)}$ exceeds $10^7$ K km s$^{-1}$ pc$^2$,
whereas there is no inter-arm GMA with such great values in $R$ and $L^{\prime}_{\rm CO(3-2)}$.
This means on-arm GMAs are relatively larger and brighter in CO($J=3-2$) than inter-arm GMAs,
suggesting inter-arm GMAs are less massive than on-arm GMAs.

\subsubsection{Virial Masses and CO Luminosity Masses}

Using physical parameters of GMAs, we estimate their masses in two different ways:
CO luminosity masses $M_{\rm CO}$ and virial masses $M_{\rm vir}$.
The former, $M_{\rm CO}$, can be calculated from CO($J=1-0$) luminosity $L^{\prime}_{\rm CO(1-0)}$
in K km s$^{-1}$ pc$^2$ unit as follows,
\begin{eqnarray}
M_{\rm CO} = 4.1 \, L^{\prime}_{\rm CO(1-0)} \,\,\,\,  M_{\odot} \,\, .
\end{eqnarray}
This is a revised version of the original formula given by \citet{bolatto2008},
who assumed the $N_{\rm H_2}$/$I_{\rm CO}$ conversion factor $X_{\rm CO} = 2.0 \times 10^{20}$ cm$^{-2}$$({\rm K \,\, km \,\, s^{-1}})^{-1}$.
Here, we adopted $X_{\rm CO} = 1.8 \times 10^{20}$ cm$^{-2}$$({\rm K \,\, km \,\, s^{-1}})^{-1}$ \citep{dame2001}
for M~83 GMAs.
However, the rotational transition of our CO data is not $J=1-0$ but $J=3-2$.
Then we have to adopt appropriate CO($J=3-2$)/CO($J=1-0$) ratios $R_{3-2/1-0}$.
Here, we assumed $R_{3-2/1-0}$ = 0.6 for the M~83 disk according to \citet{muraoka2007}.
Of course, the $R_{3-2/1-0}$ values vary locally;
\cite{muraoka2007} found that $R_{3-2/1-0}$ typically varies in the range of 0.4 to 0.8 in the M~83 disk.
Therefore, the errors of $M_{\rm CO}$ mainly originate in those of adopted $R_{3-2/1-0}$ values
and the uncertainty of $X_{\rm CO}$.
The overall error of $M_{\rm CO}$ was estimated to be about 40\%.

Then, we estimate $M_{\rm vir}$ in order to compare it with $M_{\rm CO}$.
$M_{\rm vir}$ can be calculated as follows \citep{bolatto2008}.
\begin{eqnarray}
M_{\rm vir} = 1040 \, \sigma_v^2 \, R \,\,\,\, M_{\odot}
\end{eqnarray}
where $\sigma_v$ is the velocity dispersion of GMAs in km s$^{-1}$,
and $R$ is the radius in parsec.
The error of $R$ was estimated to be 80 pc, which corresponds to
half of grid spacing of the CO($J=3-2$) map.
In addition, the error of line width was estimated to be 2.5 km s$^{-1}$,
corresponding to half of the velocity resolution.
This yields the error of $\sigma_v$ to be 1 km s$^{-1}$.
The overall error of $M_{\rm vir}$ was typically $\pm$30\%.

Figure~12 shows a relationship between $M_{\rm CO}$ and $M_{\rm vir}$ for M~83 GMAs
with that for GMAs in M~51 \citep{rand1990} and GMCs in LMC \citep{mizuno2001}. We calculate the $M_{\rm CO}$ in LMC
assuming $X_{\rm CO} = 7 \times 10^{20}$ cm$^{-2}$$({\rm K \,\, km \,\, s^{-1}})^{-1}$ \citep{fukui2008}.
CO masses of on-arm GMAs in M~83 are in the range of $10^7$ to $10^8$ $M_{\odot}$,
which is comparable with those in M~51, $10^7$ to $6 \times 10^7$ $M_{\odot}$ \citep{rand1990}.
We found that virial parameter $\alpha$, which is defined as the ratio of the virial mass
to the CO luminosity mass ($M_{\rm vir}/M_{\rm CO}$), of on-arm GMAs are almost equal to unity,
that is similar to the tendency of GMCs in LMC.
This suggests on-arm GMAs are basically in a gravitationally bound state.
On the other hand, there exist some inter-arm GMAs with $\alpha \sim$ 3 -- 10.
This suggests inter-arm GMAs are not in a gravitationally bound state,
which is different from on-arm GMAs.
Interestingly, a similar situation is also found in M~51.
\cite{rand1990} identified 6 inter-arm GMAs, and $M_{\rm vir}$ are
more than 10 times greater than $M_{\rm CO}$ for 5 of inter-arm GMAs.
On the other hand, $M_{\rm vir}$ are comparable to $M_{\rm CO}$ for most on-arm GMAs.
We emphasize that our high-quality CO($J=3-2$) data enable us to identify as many as 54 GMAs in the M~83 disk
and to confirm the difference in properties between on-arm GMAs and inter-arm GMAs more clearly.

Note that there exists an on-arm GMA with $\alpha \sim 10$ (ID num.\ 27 in table 3) in M~83.
This GMA was detected just in 4 $\sigma$ peak, so this is possibly a ``fake''.
In addition, as shown in figure~9, the systemic velocity of this GMA (507.5 km s$^{-1}$) is shifted from 
the main component (467.5 -- 477.5 km s$^{-1}$) of the spiral arm there.
Therefore, we perhaps have to refer to this GMA as an inter-arm GMA if it is not fake but ``true''.

\subsection{Origin and Nature of Inter-arm GMAs}

Finally, we discuss the origin and the nature of inter-arm GMAs in M~83.
We first briefly review the formation process of spurs, which are major substructures in inter-arm regions.
According to numerical simulations \citep{wada2004},
spur structures in inter-arm regions are formed as follows.
Molecular gas clumps in spiral arms are first formed
by the Kelvin-Helmholtz instability.
Then, some clumps fall along the spiral shocks, whereas other gases 
try to maintain nearly circular rotation.
As a result, they move away from the spiral shock towards outer spiral arms.
Since a clump has the internal gradient of its angular momentum,
the clumps gas eventually extends to inter-arm regions
as it rotates in the galactic potential.
Therefore, it is suggested that GMAs distributed in inter-arm regions are ensembles
of GMCs moving away from spiral arms in this way.
These GMCs/GMAs might be extended by shear motion,
and finally they could be destroyed and change into atomic phase.
However, as can be seen in figure~6, massive star forming regions are distributed
even in inter-arm regions, corresponding to CO($J=3-2$) spurs.
This suggests dense gas formation and massive star formation can occur within inter-arm GMAs.

Figure~13 shows a correlation between an H$\alpha$ luminosity and a virial parameter $\alpha$
for each inter-arm GMA in M~83. An anti-correlation can be seen;
GMAs with large $\alpha$ tend {\it not} to be associated with massive star forming regions.
Then, we investigate what parameter of the GMA determines $\alpha$.
Figure~14 shows a comparison between the radius $R$, velocity dispersion $\sigma_v$, and $\alpha$.
Obviously, $\sigma_v$ correlates well with $\alpha$, whereas little correlation between $R$ and $\alpha$ can be seen. 
Therefore, these plots suggest that $\alpha$ increases depending on $\sigma_v$ rather than $R$.
This dependence is also found for GMCs in the spiral arm of IC~342 \citep{hirota2008} and
at the center of the Milky way \citep{oka2001}.
Considering that a GMA is an ensemble of multiple GMCs
and the actual star formation occurs in the dense molecular cores of molecular clouds,
it is suggested that increase in the velocity dispersion of a GMC component within a GMA,
$\sigma_{v, {\rm GMC}}$, prevents the onset of star formation within the GMC and
we have observed the increase in $\sigma_{v, {\rm GMC}}$ as the increase in the overall $\sigma_v$ of the GMA
due to the inadequate spatial resolution.
Of course, the shear motion may also increase GMC-to-GMC velocity dispersions within the GMA,
contributing the increase in overall $\sigma_v$.
Therefore, we suggest that there mainly exist two kinds of inter-arm GMAs.
One is virialized ($\alpha$ $\sim$ 1) GMAs with smaller $\sigma_v$,
which are not affected by the shear motion.
Then star formation may occur within these GMAs even after moving away from spiral arms toward the inter-arm region.
The other is unvirialized GMAs with larger $\sigma_v$ and $\alpha$,
which are affected by the shear motion. Star formation process within these GMAs
might be suppressed due to the increase in the internal velocity dispersion of GMCs, $\sigma_{v, {\rm GMC}}$.

Note that since our CO observations are performed by a single dish telescope,
we cannot resolve each GMC within identified GMAs.
In order to study the relationship of $\sigma_{v, {\rm GMC}}$, $\sigma_v$, virial mass,
and star formation in more detail, high angular resolution observations using radio interferometers are necessary.

\section{Summary}

We present CO($J=3-2$) line emission observations with the Atacama Submillimeter Telescope Experiment (ASTE)
toward the $8' \times 8'$ (or 10.5 $\times$ 10.5 kpc at the distance of 4.5 Mpc) region of
the nearby barred spiral galaxy M~83 at an effective resolution of 25$^{\prime \prime}$ (550pc).
A summary of this work is as follows.

\begin{enumerate}
\item Our CO($J=3-2$) map could depict not only spiral arm structures but also
spur-like substructures extended in the inter-arm region.
We found that the spur-like substructures in CO($J=3-2$) emission are well
coincident with the distribution of massive star forming regions traced by H$\alpha$ luminosity
and Spitzer/IRAC 8 $\mu$m emission.

\item We found that the total luminosity of CO($J=3-2$) is 5.7 $\times$ $10^8$ K km s$^{-1}$ pc$^2$,
which is $\sim$ 6 times higher than the central ($r<30''$) CO($J=3-2$) luminosity, $9 \times 10^7$ K km s$^{-1}$ pc$^2$.
This result emphasizes that it is important to observe extended component in order to quantify the total luminosity
of CO even in a higher transition like $J=3-2$.

\item We found a linear and tight correlation between CO($J=3-2$) line intensity
and extinction corrected H$\alpha$ luminosity.
The index of the correlation is 0.96 $\pm$ 0.12, and its correlation coefficient $R^2$ was 0.76.
The observed slope is significantly smaller than that in the Schmidt law derived from CO($J=1-0$) emission
in the literature ($\sim$ 1.4), and it agrees well with a theoretical prediction by \cite{wada2007}. 

\item We have identified 54 CO($J=3-2$) clumps as GMAs employing the CLUMPFIND algorithm,
This is the largest GMA sample toward external galaxy disks.
We have obtained their sizes, velocity dispersions, CO($J=3-2$) luminosities, virial masses, and CO luminosity masses.
We found that on-arm GMAs are relatively larger and brighter in CO($J=3-2$) than inter-arm GMAs.

\item We found that the virial parameter $\alpha$, which is defined as the ratio of the virial mass to the CO luminosity mass,
is almost unity for GMAs in spiral arms, whereas there exist some GMAs whose $\alpha$ are 3 -- 10 in the inter-arm region.
In addition, we found that GMAs with higher $\alpha$ tend not to be associated with massive star forming regions,
while other virialized GMAs are associated with star forming regions.

\item We suggest that there mainly exist two kinds of inter-arm GMAs.
One is virialized GMAs with small $\sigma_v$,
which are not affected by the shear motion, and within which star formation may occur.
The other is unvirialized GMAs with large $\sigma_v$,
which are affected by the shear motion. Star formation process within these GMAs
might be suppressed due to the increase in the internal velocity dispersion of GMCs.

\end{enumerate} 

\vspace{0.5cm}

We would like to acknowledge the referee for invaluable comments.
We thank the members of the ASTE team for the operation and
ceaseless efforts to improve the ASTE.
Observations with ASTE were carried out 
remotely from NRO by using NTT's GEMnet2 and its partner 
R\&E (Research and Education) networks,
which are based on AccessNova 
collaboration of University of Chile,
NTT Laboratories, and National Astronomical 
Observatory of Japan.
This study was financially supported by MEXT Grant-in-Aid
for Scientific Research on Priority Areas No. 15071202.
This work is based on observations made with the Spitzer Space Telescope,
which is operated by the Jet Propulsion Laboratory, California Institute of Technology under a contract with NASA.


\clearpage



\begin{figure}
\epsscale{0.5}
\plotone{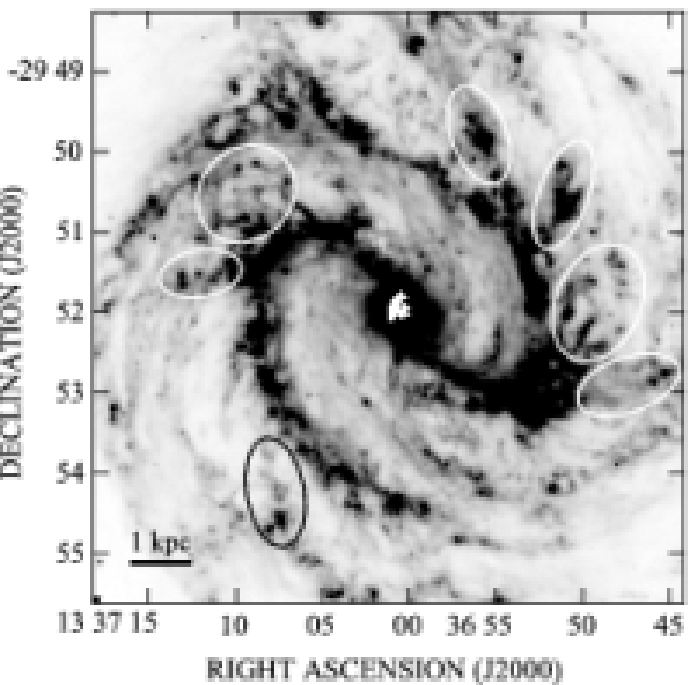}
\caption{Spitzer/IRAC 8 $\mu$m image (P40204, Robert, Kennicutt, The Local Volume Legacy Survey) of M~83.
This image depicts not only clear-cut spiral arm structures
but also a number of capillary substructures stretching from spiral arms.
These substructures are often referred to as ``spurs''.
Ellipses in the map indicate locations of spurs in 8 $\mu$m emission.
\label{fig1}}
\end{figure}

\begin{figure}
\epsscale{0.5}
\plotone{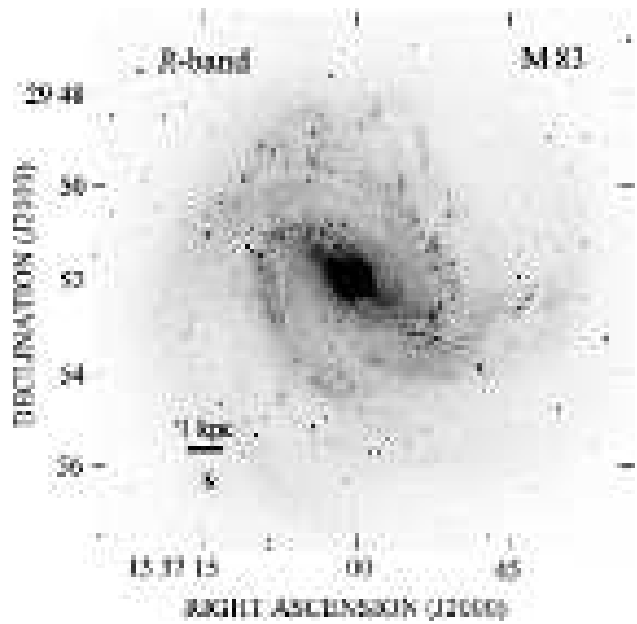}
\caption{Observed $8' \times 8'$ area (large square) using the ASTE superposed on an $R$-band image of M~83
obtained with CTIO 1.5 m \citep{meurer2006}. The whole optical disk is covered.
\label{fig2}}
\end{figure}

\begin{figure}
\epsscale{1.0}
\plotone{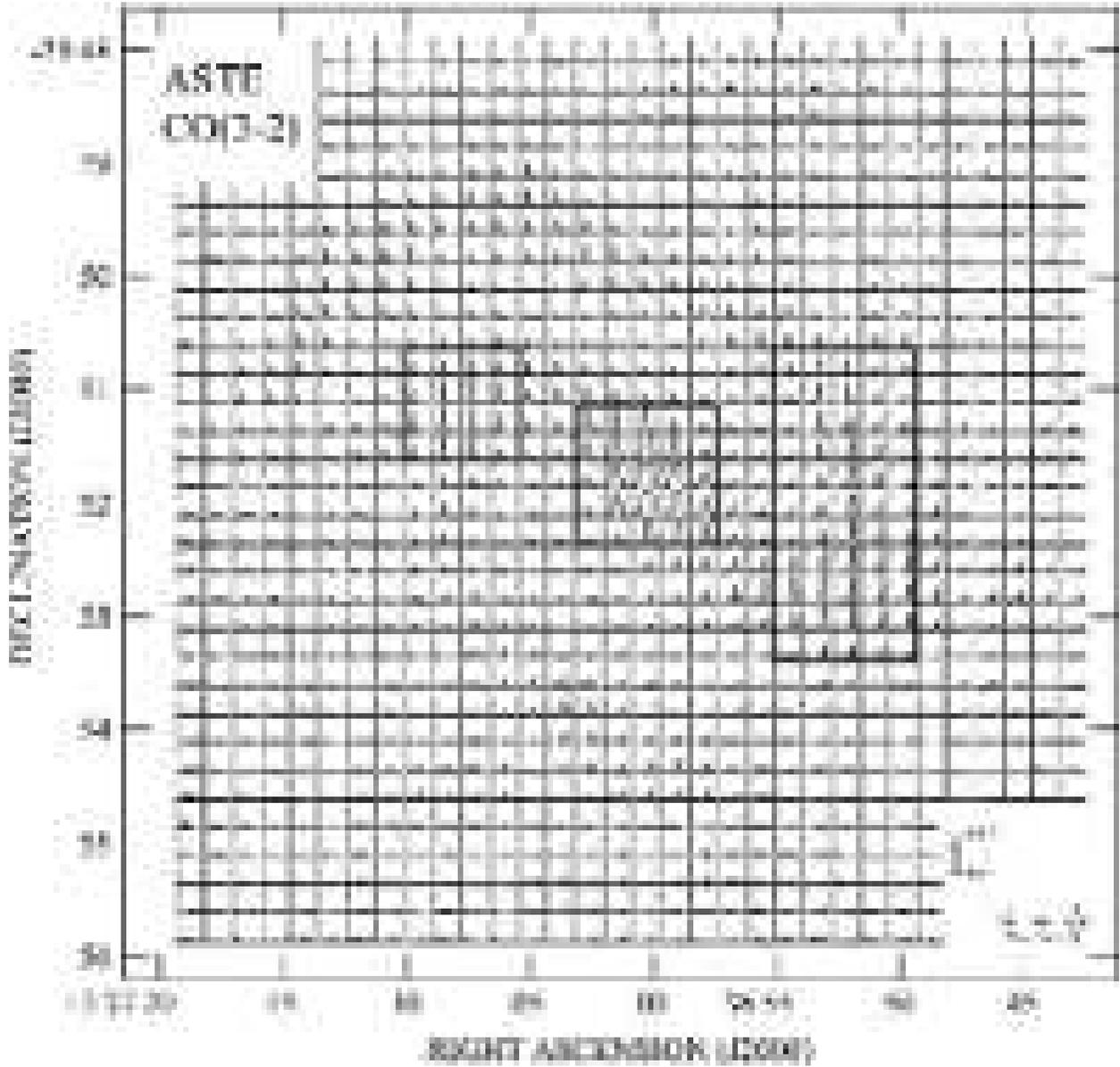}
\caption{Spectra of CO($J=3-2$) line emission in M~83. 
Although the original CO($J=3-2$) data was gridded to $7^{\prime \prime}5$,
here we show the ``thinned out'' spectra whose grid spacing is $15^{\prime \prime}$
in order to display each spectrum clearly.
The temperature scale of the spectra is indicated by the small box inserted in the lower right corner.
Significant emission is detected not only in spiral arm regions but also in inter-arm regions.
Three boxes enclosing the center, the eastern bar-end, and the western spiral arm are 
magnified and shown in figure~4.
\label{fig3}}
\end{figure}

\begin{figure}
\epsscale{0.5}
\plotone{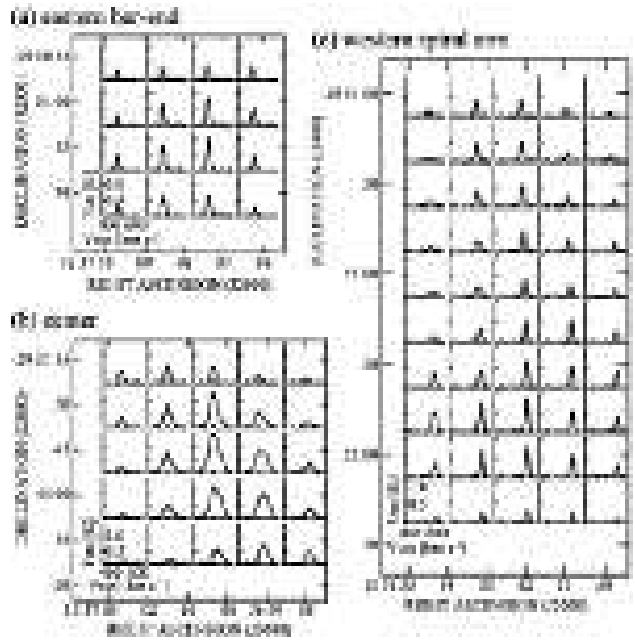}
\caption{(a) The $4 \times 4$ spectra in the eastern bar-end of M~83.
(b) The central $5 \times 5$ spectra.
(c) The $5 \times 10$ spectra in the western spiral arm. \label{fig4}}
\end{figure}

\begin{figure}
\epsscale{1.0}
\plotone{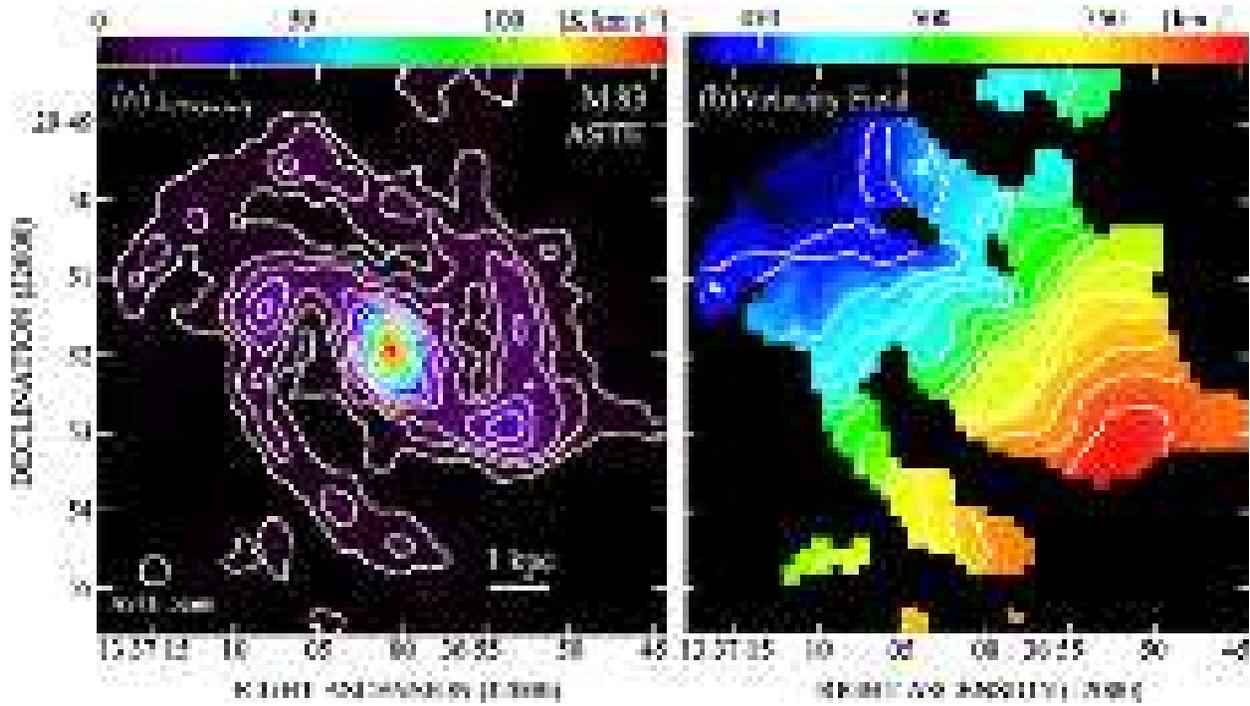}
\caption{(a) Integrated intensity map of CO($J=3-2$) line emission of M~83.
The contour levels are 3, 6, 9, 15, 21, 30, 45, 60, 90, and 135 $\sigma$, where 1 $\sigma$ = 1.0 K km s$^{-1}$.
The peak intensity is 147 K km s$^{-1}$ at the center.
(b) Velocity field measured in CO($J=3-2$) line emission. The contour levels are from 430 to 580 km s$^{-1}$
with an interval of 10 km s$^{-1}$. A warp of velocity field is clearly seen in the western spiral arm,
suggesting the existence of streaming motion.
\label{fig5}}
\end{figure}

\begin{figure}
\epsscale{1.0}
\plotone{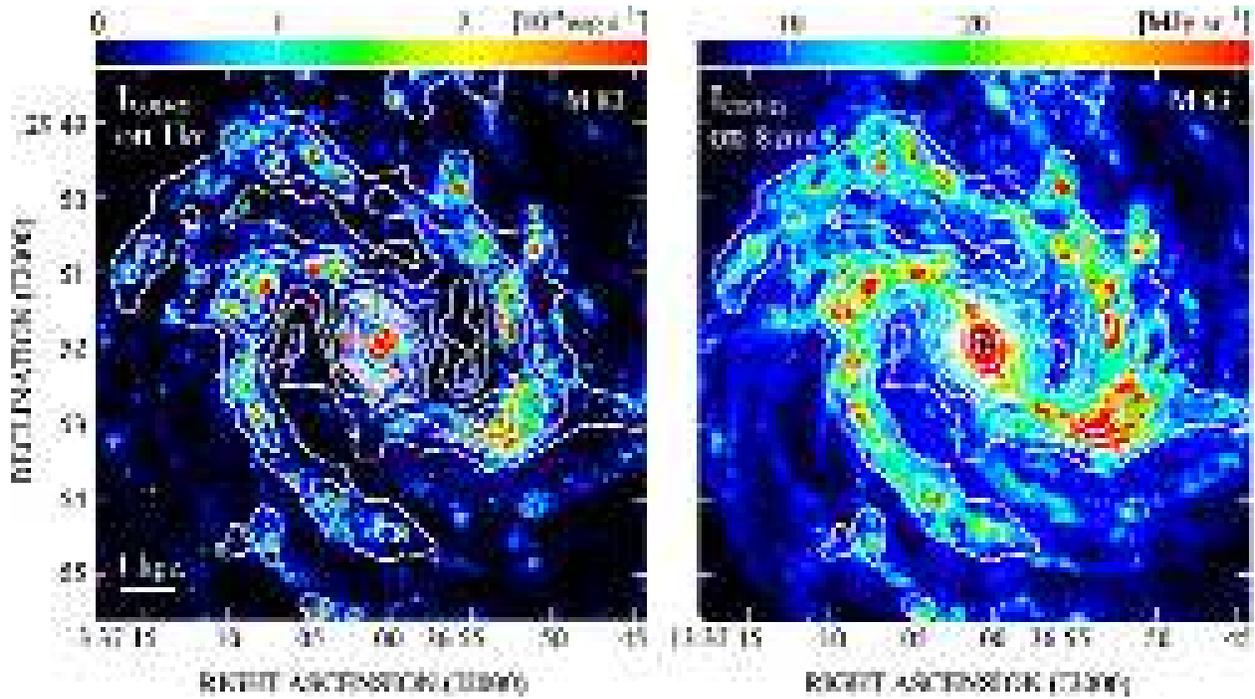}
\caption{(left) Integrated intensity map of CO($J=3-2$) line emission (contour) superposed on
continuum-subtracted H$\alpha$ luminosity map \citep{meurer2006}.
The contour levels are the same as that in figure~5.
(right) Integrated intensity map of CO($J=3-2$) line emission (contour) superposed on
IRAC 8 $\mu$m emission.
\label{fig6}}
\end{figure}

\begin{figure}
\epsscale{0.5}
\plotone{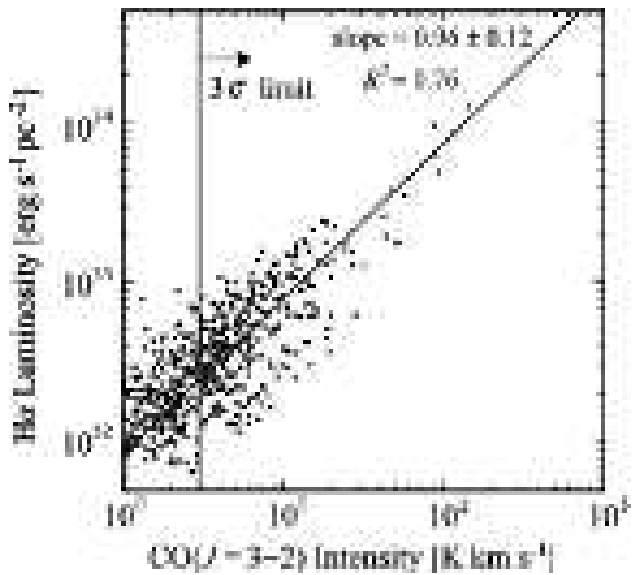}
\caption{Correlation between CO($J=3-2$) intensities and calibrated H$\alpha$ luminosities over the M~83 disk.
We have sampled each value every 15$^{\prime \prime}$ (corresponding to points where spectra are shown in figure~3).
Note that we excluded points whose CO($J=3-2$) intensities are less than 1.0 K km s$^{-1}$ in this plot.
The index of this correlation is 0.96 $\pm$ 0.12, and the correlation coefficient $R^2$ is 0.76.
\label{fig7}}
\end{figure}

\begin{figure}
\epsscale{0.5}
\plotone{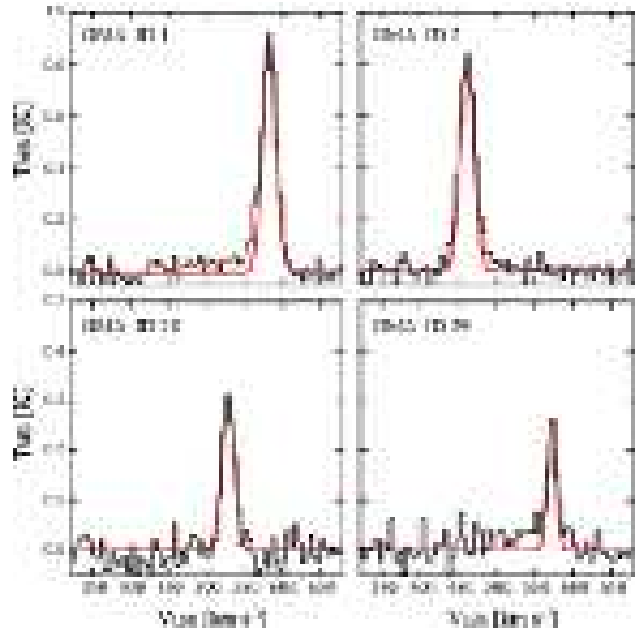}
\caption{Examples of the individual spectra of identified GMAs.
The black line indicates original spectra, and the red line indicates fitted profile by Gaussian.
The ID number of GMA is shown in each box of spectra (see table 3 and 4).
\label{fig8}}
\end{figure}

\begin{figure}
\epsscale{1}
\plotone{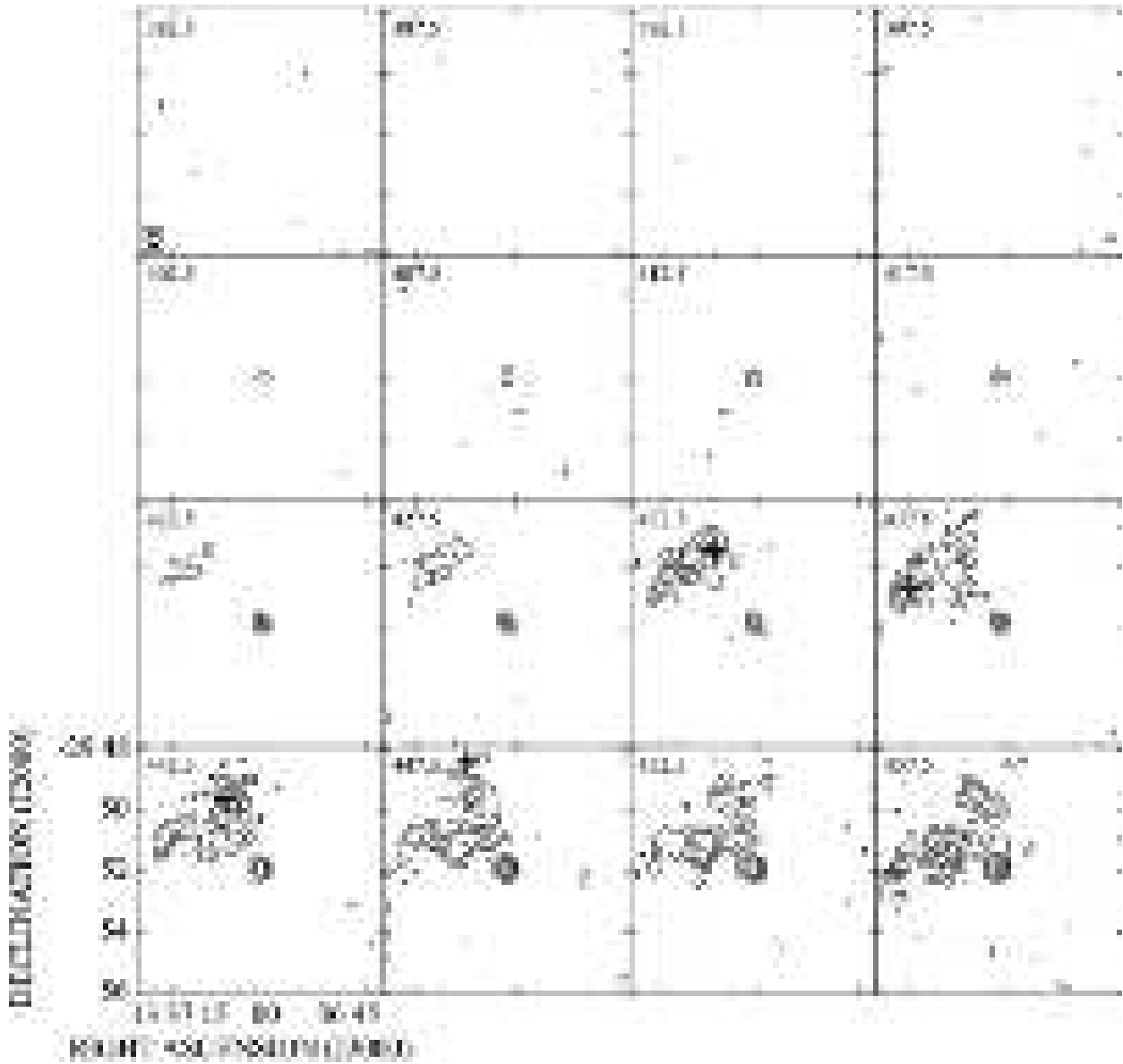}
\caption{Velocity channel maps of CO($J=3-2$) line emission in the whole optical disk of M~83.
The contour levels are 3, 6, 9, 15, 21, 30, and 45 $\sigma$, where 1 $\sigma$ = 0.028 K. 
The velocity width of each channel is 5 km s$^{-1}$, and the central velocities
($V_{\rm LSR}$ in km s$^{-1}$) are labeled in the top left corner of each map.
Crosses indicate the locations of each identified GMA,
and the number near the each cross indicate the ID of each GMA
(see table 3 and 4).
\label{fig9a}}
\end{figure}

\setcounter{figure}{8}
\begin{figure}
\epsscale{1}
\plotone{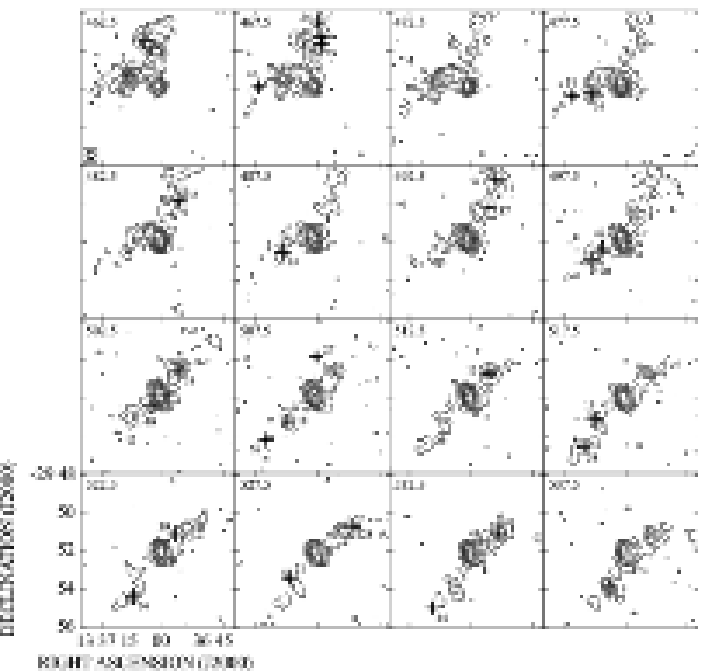}
\caption{continued.
\label{fig9b}}
\end{figure}

\setcounter{figure}{8}
\begin{figure}
\epsscale{1}
\plotone{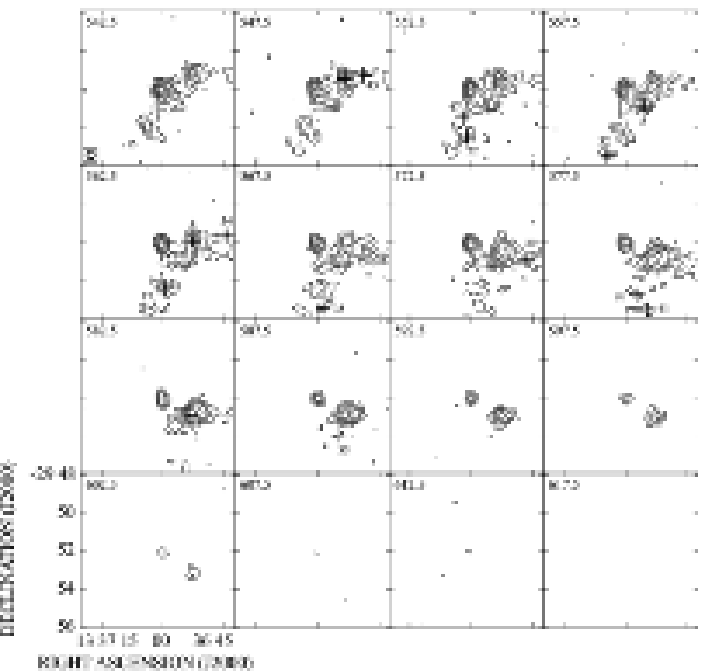}
\caption{continued.
\label{fig9c}}
\end{figure}

\begin{figure}
\epsscale{0.5}
\plotone{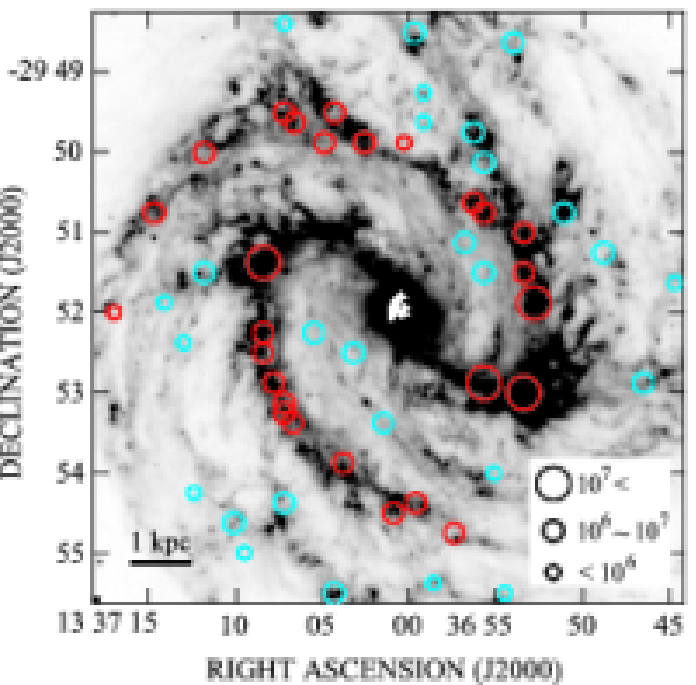}
\caption{GMA distribution in the M~83 disk
superposed on the Spitzer/IRAC 8 $\mu$m map.
Colors of circles indicates each region;
red indicates on-arm GMAs,
and blue indicates inter-arm GMAs.
In addition, the size of the circles reflect CO($J=3-2$) luminosity of each GMA;
large circles indicate GMAs whose CO($J=3-2$) luminosities exceed $10^7$ K km s$^{-1}$ pc$^2$,
intermediate circles indicate GMAs whose CO($J=3-2$) luminosities are in the range of $10^6$ K km s$^{-1}$ pc$^2$ to
$10^7$ K km s$^{-1}$ pc$^2$,
and small circles indicate GMAs whose CO($J=3-2$) luminosity are falling below $10^6$ K km s$^{-1}$ pc$^2$, respectively.
The correspondence between CO($J=3-2$) luminosities and circle sizes is shown in the lower right corner.
\label{fig10}}
\end{figure}

\begin{figure}
\epsscale{0.5}
\plotone{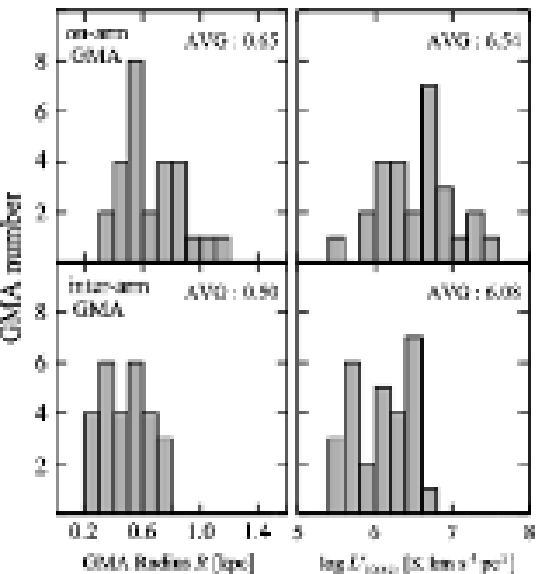}
\caption{Histograms of radius $R$, and $L^{\prime}_{\rm CO(3-2)}$ for on-arm GMAs (top) and inter-arm GMAs (bottom).
The caption ``AVG'' means an arithmetic average for $R$,
and means a geometrical average for $L^{\prime}_{\rm CO(3-2)}$.
On-arm GMAs are relatively larger and brighter in CO($J=3-2$) than inter-arm GMAs.
\label{fig11}}
\end{figure}

\begin{figure}
\epsscale{0.5}
\plotone{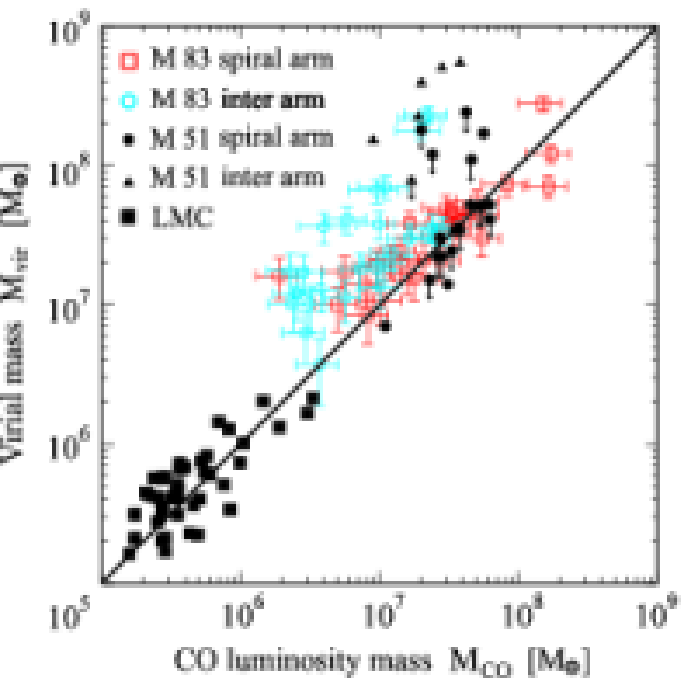}
\caption{Comparison between virial mass and CO luminosity mass for GMAs in M~83 (this work)
and M~51 \citep{rand1990}, and GMCs in LMC \citep{mizuno2001}.
Most of on-arm GMAs seem virialized, whereas there exists some inter-arm GMAs whose virial masses
are 3 -- 10 times higher than their CO luminosity masses.
\label{fig12}}
\end{figure}

\begin{figure}
\epsscale{0.5}
\plotone{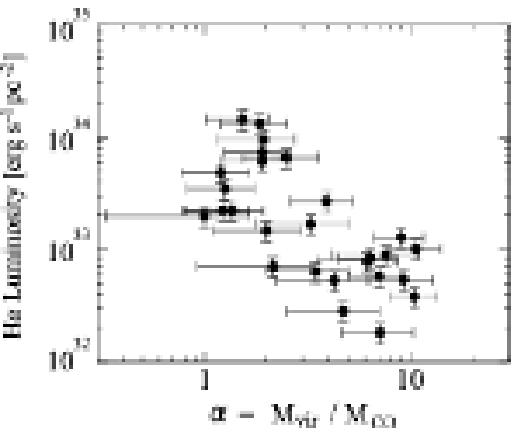}
\caption{Correlation between H${\alpha}$ luminosity and the virial parameter $\alpha$defined as
the ratio of virial mass per CO luminosity mass ($M_{\rm vir}$/$M_{\rm CO}$) for each inter-arm GMA.
\label{fig13}}
\end{figure}

\begin{figure}
\epsscale{1.0}
\plotone{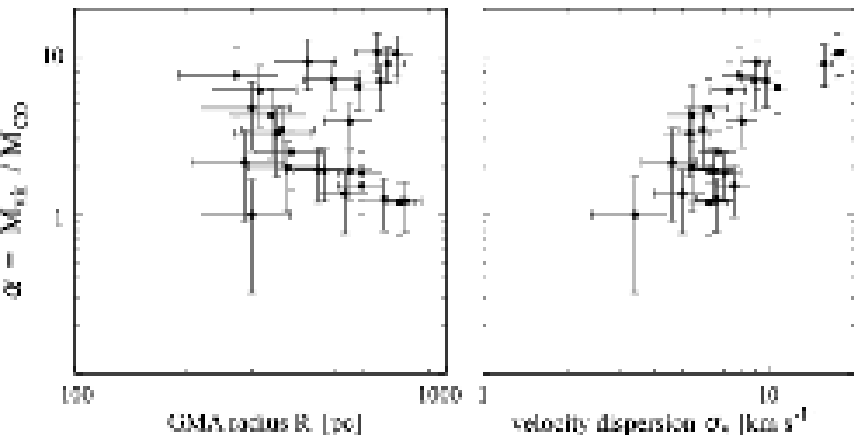}
\caption{(left) Comparison between the radius $R$ and the virial parameter $\alpha$
for inter-arm GMAs. Little correlation can be seen.
(right) Comparison between the velocity dispersion $\sigma_{v}$ and $\alpha$.
A tight correlation can be found.
\label{fig14}}
\end{figure}

\clearpage









\clearpage


\begin{table}
\begin{center}
Table~1.\hspace{4pt}Observation parameters of M~83 in CO($J=3-2$) line.\\[1mm]
\begin{tabular}{ll}
\hline \hline \\[-6mm]
Telescope & ASTE \\[-2mm]
\,\,\,\, diameter & \,\,\,\,10 m\\[-2mm]
\,\,\,\, beam size & \,\,\,\,$22^{\prime \prime}$\\[-2mm]
\,\,\,\, main-beam efficiency$^{\ast}$ & \,\,\,\,0.57 $\pm$ 0.06 $\pm$ 0.03\\
Receiver & CATS345\\[-2mm]
\,\,\,\,receiver type & \,\,\,\,2SB$^{\ast \ast}$\\[-2mm]
\,\,\,\,receiver noise temperature & \,\,\,\,$\sim$ 100 K \\[-2mm]
\,\,\,\,image rejection ratio & \,\,\,\,$\sim$ 10 dB\\
Spectrometer & MAC \\[-2mm]
\,\,\,\,channel number & \,\,\,\,1024 \\[-2mm]
\,\,\,\,frequency coverage & \,\,\,\,512 MHz\\[-2mm]
\,\,\,\,velocity coverage & \,\,\,\,440 ${\rm km}$ ${\rm s}^{-1}$\\
Map description & \\[-2mm]
\,\,\,\,mapping mode & \,\,\,\,On-the-fly\\[-2mm]
\,\,\,\,field coverage & \,\,\,\,$8^{\prime} \times 8^{\prime}$ (10.5 $\times$ 10.5 kpc)\\[-2mm]
\,\,\,\,rms noise level ($\Delta v$=5 km s$^{-1}$)& \,\,\,\,25 -- 30 mK (in $T_{\rm MB}$ scale)\\
\hline \\[-2mm]
\end{tabular}\\
{\footnotesize
$^{\ast}\hspace{3pt}$For the main-beam efficiency, the first error indicates systematic error
and the second, random error.\\
$^{\ast \ast}$2SB means side band separating mixer receiver.
}
\end{center}
\end{table}

\clearpage


\begin{table}
\begin{center}
Table~2.\hspace{4pt}Map parameters and physical properties of M~83.\\[1mm]
\begin{tabular}{lll}
\hline \hline \\[-3mm]
r.m.s. noise level & 1.0 K km s$^{-1}$\\
effective spatial resolution & $\sim$ 25$^{\prime \prime}$\\
Peak intensity & 146.7 K km s$^{-1}$\\
$L^{\prime}_{\rm CO(3-2)}$ & 5.7 $\times$ $10^8$ K km s$^{-1}$ pc$^2$\\
\hline \\[-2mm]
\end{tabular}\\
{\footnotesize
}
\end{center}
\end{table}

\clearpage


\begin{table}
\begin{center}
Table~3.\hspace{4pt}Identified {\it on-arm} GMA properties of M~83.\\[1mm]
\begin{tabular}{ccccccccccc}
\hline \hline \\[-6mm]
ID       & R.A.    & Decl. & $D$ & $V_{\rm LSR}$ & $T_{\rm peak}$ & $R$ & $\sigma_v$ & $L^{\prime}_{\rm CO(3-2)}$ & $M_{\rm CO}$ & $M_{\rm vir}$ \\[-2mm]
(1)      & (2)     & (3)   & (4) & (5)           & (6)            & (7) & (8)        & (9)                        & (10)         & (11)          \\
\hline \\[-6mm]
      1   & 13 36 53.4	& -29 53 00.4	& 2.6	& 582.5	& 0.91	& 960	& \,\,\,8.4	&     24.1 &           162.4	& \,\,\,70.9 \\[-3mm]
      2   & 13 37 08.4	& -29 51 22.9	& 2.3	& 462.5	& 0.83	& 1030	&      10.7	&     25.2 &           170.1	& 122.4 \\[-3mm]
      3   & 13 36 53.4	& -29 51 30.4	& 2.2	& 547.5	& 0.61	& 770	& \,\,\,7.5	&\,\,\,7.8 &      \,\,\,52.6	& \,\,\,44.7 \\[-3mm]
      4   & 13 36 52.8	& -29 51 52.9	& 2.3	& 562.5	& 0.54	& 870	& \,\,\,9.0	&     12.1 &      \,\,\,81.6	& \,\,\,73.3 \\[-3mm]
      5   & 13 36 55.7	& -29 52 52.9	& 1.9	& 557.5	& 0.51	& 1120	&      15.4	&     22.1 &           149.4	& 279.0 \\[-3mm]
      6   & 13 36 53.4	& -29 51 00.4	& 2.5	& 532.5	& 0.48	& 540	& \,\,\,7.9	&\,\,\,5.1 &      \,\,\,34.3	& \,\,\,34.6 \\[-3mm]
      7   & 13 37 02.6	& -29 49 52.9	& 2.8	& 462.5	& 0.40	& 870	& \,\,\,7.9	&\,\,\,7.5 &      \,\,\,50.4	& \,\,\,55.8 \\[-3mm]
      8   & 13 37 03.8	& -29 53 52.9	& 2.7	& 537.5	& 0.38	& 710	& \,\,\,7.7	&\,\,\,4.8 &      \,\,\,32.2	& \,\,\,44.0 \\[-3mm]
      9   & 13 37 11.8	& -29 50 00.4	& 4.0	& 432.5	& 0.36	& 890	& \,\,\,5.7	&\,\,\,8.0 &      \,\,\,53.9	& \,\,\,29.9 \\[-3mm]
     10   & 13 37 06.6	& -29 49 37.9	& 3.5	& 442.5	& 0.35	& 570	& \,\,\,6.0	&\,\,\,3.1 &      \,\,\,20.7	& \,\,\,21.5 \\[-3mm]
     11   & 13 37 07.2	& -29 49 30.4	& 3.7	& 432.5	& 0.34	& 740	& \,\,\,5.2	&\,\,\,4.0 &      \,\,\,27.2	& \,\,\,20.8 \\[-3mm]
     12   & 13 37 04.3	& -29 49 30.4	& 3.4	& 462.5	& 0.34	& 610	& \,\,\,7.0	&\,\,\,4.1 &      \,\,\,27.9	& \,\,\,31.3 \\[-3mm]
     13   & 13 36 55.7	& -29 50 45.4	& 2.2	& 512.5	& 0.33	& 560	& \,\,\,5.2	&\,\,\,2.6 &      \,\,\,17.6	& \,\,\,15.5 \\[-3mm]
     14   & 13 36 56.3	& -29 50 37.9	& 2.2	& 502.5	& 0.33	& 760	& \,\,\,7.7	&\,\,\,4.9 &      \,\,\,32.9	& \,\,\,46.4 \\[-3mm]
     15   & 13 37 14.7	& -29 50 45.4	& 4.2	& 437.5	& 0.32	& 810	& \,\,\,7.0	&\,\,\,6.2 &      \,\,\,41.7	& \,\,\,41.7 \\[-3mm]
     16   & 13 37 08.4	& -29 52 15.5	& 2.2	& 477.5	& 0.29	& 570	& \,\,\,9.1	&\,\,\,4.5 &      \,\,\,30.1	& \,\,\,49.8 \\[-3mm]
     17   & 13 36 59.6	& -29 54 22.9	& 3.2	& 562.5	& 0.27	& 560	& \,\,\,5.3	&\,\,\,2.3 &      \,\,\,15.7	& \,\,\,16.6 \\[-3mm]
     18   & 13 37 07.2	& -29 53 15.4	& 2.5	& 507.5	& 0.25	& 500	& \,\,\,4.6	&\,\,\,1.5 &      \,\,\,10.0	& \,\,\,10.8 \\[-3mm]
     19   & 13 37 07.2	& -29 53 07.9	& 2.4	& 517.5	& 0.23	& 480	& \,\,\,4.1	&\,\,\,1.2 & \,\,\,\,\,\,8.1	& \,\,\,\,\,\,8.4 \\[-3mm]
     20   & 13 37 07.8	& -29 52 52.9	& 2.3	& 497.5	& 0.23	& 510	& \,\,\,6.3	&\,\,\,1.8 &      \,\,\,11.9	& \,\,\,21.0 \\[-3mm]
     21   & 13 37 06.7	& -29 53 22.9	& 2.5	& 527.5	& 0.20	& 380	& \,\,\,6.0	&\,\,\,1.1 & \,\,\,\,\,\,7.7	& \,\,\,14.2 \\[-3mm]
     22   & 13 37 00.9	& -29 54 30.4	& 3.4	& 552.5	& 0.20	& 570	& \,\,\,8.0	&\,\,\,2.4 &      \,\,\,16.4	& \,\,\,38.4 \\[-3mm]
     23   & 13 37 08.4	& -29 52 30.4	& 2.3	& 487.5	& 0.20	& 430	& \,\,\,6.5	&\,\,\,1.4 & \,\,\,\,\,\,9.3	& \,\,\,19.0 \\[-3mm]
     24   & 13 37 04.9	& -29 49 52.9	& 2.9	& 442.5	& 0.20	& 630	& \,\,\,6.1	&\,\,\,2.1 &      \,\,\,14.0	& \,\,\,24.4 \\[-3mm]
     25   & 13 37 17.0	& -29 52 00.3	& 4.6	& 457.5	& 0.16	& 410	& \,\,\,6.3	& \,\,\,0.8& \,\,\,\,\,\,5.6	& \,\,\,17.3 \\[-3mm]
     26   & 13 36 57.4	& -29 54 45.4	& 3.8	& 577.5	& 0.15	& 410	& \,\,\,4.9	&\,\,\,0.7 & \,\,\,\,\,\,4.9	& \,\,\,10.0 \\[-3mm]
     27   & 13 37 00.3	& -29 49 52.9	& 2.7	& 507.5	& 0.11	& 300	& \,\,\,7.2	& \,\,\,0.3& \,\,\,\,\,\,1.9	& \,\,\,16.3 \\
\hline \\[-2mm]
\multicolumn{11}{@{}l@{}}{\hbox to 0pt{\parbox{160mm}{\footnotesize
Notes. (1): ID number of each GMA.
(2)-(3): Positions of the peak CO($J=3-2$) intensity of each GMA in equatorial coordinates (J2000).
Units of right ascension are hours, minutes, and seconds,
and units of declination are degrees, arcminutes, and arcseconds.
(4): Deprojected distance from the galaxy center in kpc.
(5): $V_{\rm LSR}$ of each GMA in km s$^{-1}$.
(6): Peak CO($J=3-2$) intensity of each GMA in K.
(7): Deconvolved GMA radius in pc.
(8): Velocity dispersion in km s$^{-1}$.
(9): Derived CO($J=3-2$) luminosity in $10^6$ K km s$^{-1}$ pc$^2$.
(10): GMA mass derived from CO luminosity in $10^6$ $M_{\odot}$.
(11): Virial mass in $10^6$ $M_{\odot}$.}\hss}}
\end{tabular}\\
\end{center}
\end{table}

\clearpage


\begin{table}
\begin{center}
Table~4.\hspace{4pt}Identified {\it inter-arm} GMA properties of M~83.\\[1mm]
\begin{tabular}{ccccccccccc}
\hline \hline \\[-6mm]
ID       & R.A.    & Decl. & $D$ & $V_{\rm LSR}$ & $T_{\rm peak}$ & $R$ & $\sigma_v$ & $L^{\prime}_{\rm CO(3-2)}$ & $M_{\rm CO}$ & $M_{\rm vir}$ \\[-2mm]
(1)      & (2)     & (3)   & (4) & (5)           & (6)            & (7) & (8)        & (9)                        & (10)         & (11)          \\
\hline \\[-6mm]
   28      & 13 36 51.1	& -29 50 45.4	& 3.2	& 527.5	& 0.31	& 590	& \,\,\,7.6	& \,\,\,3.5&      \,\,\,23.6	& \,\,\,36.0 \\[-3mm]
   29      & 13 36 46.5	& -29 52 52.9	& 4.3	& 572.5	& 0.26	& 780	& \,\,\,6.5	& \,\,\,4.2&      \,\,\,28.3	& \,\,\,34.5 \\[-3mm]
   30      & 13 37 11.8	& -29 51 30.4	& 3.2	& 457.5	& 0.26	& 680	& \,\,\,6.6	& \,\,\,3.7&      \,\,\,24.8	& \,\,\,31.2 \\[-3mm]
   31      & 13 36 54.0	& -29 48 37.9	& 4.8	& 492.5	& 0.26	& 750	& \,\,\,6.2	& \,\,\,3.7&      \,\,\,24.6	& \,\,\,29.7 \\[-3mm]
   32      & 13 37 04.3	& -29 55 30.4	& 4.8	& 557.5	& 0.22	& 450	& \,\,\,6.2	& \,\,\,1.4& \,\,\,\,\,\,9.5	& \,\,\,18.3 \\[-3mm]
   33      & 13 36 56.3	& -29 49 45.4	& 3.2	& 482.5	& 0.21	& 590	& \,\,\,7.0	& \,\,\,2.4&      \,\,\,16.1	& \,\,\,29.8 \\[-3mm]
   34      & 13 36 59.7	& -29 48 30.4	& 4.5	& 467.5	& 0.20	& 470	& \,\,\,6.5	& \,\,\,1.6&      \,\,\,10.7	& \,\,\,20.5 \\[-3mm]
   35      & 13 36 58.6	& -29 55 22.9	& 4.6	& 567.5	& 0.19	& 370	& \,\,\,5.4	& \,\,\,0.8& \,\,\,\,\,\,5.6	& \,\,\,11.3 \\[-3mm]
   36      & 13 37 07.2	& -29 54 22.9	& 3.7	& 522.5	& 0.18	& 380	& \,\,\,6.7	& \,\,\,1.1& \,\,\,\,\,\,7.3	& \,\,\,18.1 \\[-3mm]
   37      & 13 36 55.7	& -29 50 07.9	& 2.8	& 492.5	& 0.18	& 550	& \,\,\,6.4	& \,\,\,1.8&      \,\,\,12.4	& \,\,\,23.6 \\[-3mm]
   38      & 13 37 10.1	& -29 54 37.9	& 4.4	& 517.5	& 0.18	& 530	& \,\,\,5.0	& \,\,\,1.5&      \,\,\,10.0	& \,\,\,13.6 \\[-3mm]
   39      & 13 36 56.8	& -29 51 07.9	& 1.6	& 522.5	& 0.17	& 700	&      15.7	& \,\,\,2.9&      \,\,\,19.7	&      178.4 \\[-3mm]
   40      & 13 37 03.2	& -29 52 30.4	& 1.0	& 502.5	& 0.17	& 650	&      17.9	& \,\,\,3.0&      \,\,\,20.5	&      216.9 \\[-3mm]
   41      & 13 36 54.5	& -29 55 30.4	& 5.0	& 577.5	& 0.17	& 300	& \,\,\,3.4	& \,\,\,0.5& \,\,\,\,\,\,3.7	& \,\,\,\,\,\,3.7 \\[-3mm]
   42      & 13 37 05.5	& -29 52 15.4	& 1.4	& 497.5	& 0.16	& 580	&      10.8	& \,\,\,1.6&      \,\,\,11.0	& \,\,\,70.1 \\[-3mm]
   43      & 13 36 55.7	& -29 51 30.4	& 1.6	& 532.5	& 0.15	& 740	&      17.4	& \,\,\,3.3&      \,\,\,22.3	&      231.3 \\[-3mm]
   44      & 13 37 09.5	& -29 55 00.4	& 4.7	& 532.5	& 0.15	& 490	& \,\,\,9.0	& \,\,\,0.9& \,\,\,\,\,\,5.8	& \,\,\,41.2 \\[-3mm]
   45      & 13 36 48.8	& -29 51 15.4	& 3.6	& 547.5	& 0.14	& 550	& \,\,\,8.1	& \,\,\,1.4& \,\,\,\,\,\,9.8	& \,\,\,38.0 \\[-3mm]
   46      & 13 37 14.1	& -29 51 52.9	& 3.8	& 467.5	& 0.13	& 360	& \,\,\,5.9	& \,\,\,0.6& \,\,\,\,\,\,3.8	& \,\,\,13.0 \\[-3mm]
   47      & 13 37 01.5	& -29 53 22.9	& 1.9	& 552.5	& 0.12	& 660	& \,\,\,9.8	& \,\,\,1.4& \,\,\,\,\,\,9.5	& \,\,\,66.4 \\[-3mm]
   48      & 13 36 59.2	& -29 49 15.4	& 3.6	& 467.5	& 0.12	& 350	& \,\,\,5.3	& \,\,\,0.5& \,\,\,\,\,\,3.1	& \,\,\,10.1 \\[-3mm]
   49      & 13 36 59.2	& -29 49 37.9	& 3.1	& 467.5	& 0.12	& 290	& \,\,\,4.6	& \,\,\,0.4& \,\,\,\,\,\,2.9	& \,\,\,\,\,\,6.3 \\[-3mm]
   50      & 13 37 07.2	& -29 48 23.8	& 5.0	& 447.5	& 0.12	& 310	& \,\,\,7.3	& \,\,\,0.4& \,\,\,\,\,\,2.8	& \,\,\,17.2 \\[-3mm]
   51      & 13 37 13.0	& -29 52 22.9	& 3.5	& 477.5	& 0.12	& 340	& \,\,\,5.4	& \,\,\,0.4& \,\,\,\,\,\,2.4	& \,\,\,10.1 \\[-3mm]
   52      & 13 37 12.4	& -29 54 15.4	& 4.5	& 507.5	& 0.11	& 300	& \,\,\,6.2	& \,\,\,0.4& \,\,\,\,\,\,2.5	& \,\,\,12.0 \\[-3mm]
   53      & 13 36 55.1	& -29 54 00.4	& 3.2	& 587.5	& 0.11	& 270	& \,\,\,7.9	& \,\,\,0.3& \,\,\,\,\,\,2.3	& \,\,\,17.4 \\[-3mm]
   54      & 13 36 44.7	& -29 51 37.8	& 4.6	& 562.5	& 0.11	& 420	& \,\,\,9.1	& \,\,\,0.6& \,\,\,\,\,\,3.9	& \,\,\,36.4 \\
\hline \\[-2mm]
\multicolumn{11}{@{}l@{}}{\hbox to 0pt{\parbox{160mm}{\footnotesize
Notes. (1): ID number of each GMA.
(2)-(3): Positions of the peak CO($J=3-2$) intensity of each GMA in equatorial coordinates (J2000).
Units of right ascension are hours, minutes, and seconds,
and units of declination are degrees, arcminutes, and arcseconds.
(4): Deprojected distance from the galaxy center in kpc.
(5): $V_{\rm LSR}$ of each GMA in km s$^{-1}$.
(6): Peak CO($J=3-2$) intensity of each GMA in K.
(7): Deconvolved GMA radius in pc.
(8): Velocity dispersion in km s$^{-1}$.
(9): Derived CO($J=3-2$) luminosity in $10^6$ K km s$^{-1}$ pc$^2$.
(10): GMA mass derived from CO luminosity in $10^6$ $M_{\odot}$.
(11): Virial mass in $10^6$ $M_{\odot}$.}\hss}}
\end{tabular}\\
\end{center}
\end{table}

\end{document}